\documentclass[epj]{svjour}
\usepackage{color}
\usepackage{ulem}
\usepackage{graphicx}
\usepackage{dcolumn}
\usepackage[T1]{fontenc}
\usepackage{amsmath}
\usepackage{amssymb}
\usepackage[english]{babel}
\usepackage{graphics}
\usepackage{cite}

\def\MSbar{\overline{\text{MS}}}

\def\GeV{\text{ GeV}}
\def\MeV{\text{ MeV}}

\begin{document}
\title{Spin structure of the nucleon: \\QCD evolution, lattice results and models}
\subtitle{\\}
\author{M.~Altenbuchinger\inst{1}\thanks{altenb@ph.tum.de} \and Ph.~H\"agler\inst{1,2}\thanks{phaegler@ph.tum.de} \and W.~Weise\inst{1}\thanks{weise@ph.tum.de} \and E.~M.~Henley\inst{3}}
 \institute{Physik Department, Technische Universit\"at M\"unchen, D-85747 Garching, Germany \and Institut f\"ur Theoretische Physik, Universit\"at Regensburg, D-93040 Regensburg, Germany\and Dept. of Physics, University of Washington, Seattle, WA 98195-1560, USA}
\date{Received: date / Revised version: date}

\abstract{
The question how the spin of the nucleon is distributed among its quark
and gluon constituents is still a subject of intense investigations. Lattice QCD has progressed to provide information about spin fractions and orbital angular momentum contributions for up- and down-quarks in the proton, at a typical scale $\mu^2\sim 4\,\rm{GeV}^2$. On the other hand, chiral quark models have traditionally been used for orientation at low momentum scales. In the comparison of such model calculations with experiment or lattice QCD, fixing the model scale and the treatment of scale evolution are essential. In this paper, we present a refined model calculation and a QCD evolution of lattice results up to next-to-next-to-leading order. We compare this approach with the Myhrer-Thomas scenario for resolving the proton spin puzzle. \\
}

\maketitle

\section{Introduction}
How is the total spin  $1/2$ of the nucleon distributed among its quark and gluon constituents? This question has been intensely discussed ever since the EMC experiment presented first results for the spin asymmetry in polarized muon proton scattering in 1987 \cite{Ashman:1987hv}. This measurement indicated that only about 15\% or less of the nucleon spin is built up by quark spins, although with sizeable statistical and systematic uncertainties. More recent measurements of HERMES and COMPASS \cite{Airapetian:2007mh,Alexakhin:2006vx} and their QCD analysis \cite{Blumlein:2010rn,deFlorian:2009vb,Leader:2010rb} showed that the nucleon receives still only about one third of its spin from quark spins: 
\begin{eqnarray}
\Delta \Sigma_{\text{\tiny{HERMES}}}(5\,\mbox{GeV}^2)&&=\nonumber\\
0.330\pm0.011_\text{\tiny{theo.}}&\pm& 0.025_\text{\tiny{exp.}}\pm 0.028_\text{\tiny{evol.}},
\label{Hermes}
\end{eqnarray} 
determined at a scale $\mu^2=5\,\rm{GeV}^2$. This is in stark contrast to naive model calculations, as for example in the non-relativistic quark model that suggests $\Delta \Sigma=1$. Relativistic effects reduce $\Delta \Sigma$ to about two thirds, still far too large in comparison with Eq.~(\ref{Hermes}).
Myhrer and Thomas proposed in \cite{Myhrer:2007cf,Thomas:2008ga,Thomas:2009gk,Thomas:2008bd} that $\Delta \Sigma$ could be further reduced by including pion cloud contributions and corrections from one gluon exchanges. With such corrections they end up with a result for $\Delta\Sigma$ that is consistent with experiment. The missing $\approx 60-70\%$ of the nucleon spin reappear entirely as orbital angular momentum of up and down quarks. On the other hand, $L_{u+d}$, appears to be in strong contrast to lattice calculations \cite{Hagler:2003jd,Gockeler:2003jf,Hagler:2007xi}
 where the orbital angular momentum contribution $L_{u+d}$ comes out close to zero \cite{Hagler:2007xi}. To explain this difference, Thomas \cite{Thomas:2008ga} proposed to consider the renormalization scale \mbox{($\mu$-)}dependence of the quantities appearing in the nucleon spin sum rule \cite{Jaffe:1989jz} 
\begin{equation}
\label{sumrule}
\frac{1}{2}\Delta \Sigma+L_q+L_g+\Delta G=\frac{1}{2},
\end{equation}
defined by the following expectation values taken for a spin-up state of the proton, $|P+\rangle$:
\begin{eqnarray}
\label{spin operators}
\Delta \Sigma&=&\langle P +| \int d^3x \bar \psi \gamma^3 \gamma_5 \psi |P+\rangle,\nonumber\\
\Delta G&=&\langle P +| \int d^3x (E^1A^2-E^2A^1) |P+\rangle,\nonumber\\
L_q&=&\langle P +| i\int d^3x\psi^\dagger (x^1 \partial^2-x^2 \partial^1)\psi |P+\rangle,\nonumber\\
L_g&=&\langle P +| \int d^3x E^i(x^2 \partial^1-x^1 \partial^2)A^i |P+\rangle.
\label{eq-spincontrib}
\end{eqnarray}
Here $\psi$ is the quark field, $E^i$ and $A^\mu$ are the gluon electric field and gauge potential. A sum over quark flavors is implicit in the definition of the flavor singlet quantities in Eq.~(\ref{spin operators}). Contributions in the non-singlet sector will be denoted by $\Delta \Sigma_{u-d}$, $L_{u-d}$ etc.~. The $L_g$ is the orbital angular momentum contribution from gluons and $\Delta G$ is the gluon spin part. 
It is important to note that $L_q$, $L_g$ and $\Delta G$ in Eq.~(\ref{eq-spincontrib}) are 
not explicitly gauge invariant. 
A manifestly gauge invariant decomposition and its relation to moments of generalized parton distributions
was presented by Ji in \cite{Ji:1995cu,Ji:1996ek}:
\begin{equation}
\label{spin operators2}
\frac{1}{2}\Delta \Sigma+L^{\text{GI}}_{q}+J^{\text{GI}}_g=\frac{1}{2},
\end{equation}
where $\Delta \Sigma$ is given as before, $L^{\text{GI}}_{q}$ is obtained from $L_q$ replacing $\partial^\mu$ by the gauge-covariant derivative, $\partial^\mu \rightarrow D^\mu$,
and the total gluon angular momentum is defined as 
\begin{equation}
\label{spin operators3}
J^{\text{GI}}_g=\langle P +|\int d^3x[\vec x \times (\vec E \times \vec B)]_3|P+\rangle\, .
\end{equation}
Using a leading order QCD evolution of the spin contributions from the low, hadronic model scale
to the higher scale of the lattice results, it was shown in Ref.~\cite{Thomas:2008ga} that 
it is possible to find at least a qualitative agreement with the lattice data.

With these previous achievements in mind, the purpose of the present work is twofold:
first, we extend the QCD evolution to next-to-leading (NLO) and next-to-next-to-leading (NNLO) order and perform a backwards evolution starting from lattice results. This approach has the advantage that the 
scale dependence of the spin contributions is rather weak at the higher scale of lattice results, and that the
extrapolation therefore does not suffer from the uncertainty of the slope in $\mu$ at low scales.
Most importantly, proceeding in this way we do not have to fix the model scale a priori, which is generically difficult,
but have the possibility to compare model results over a wider range of low scales with the downward-evolved lattice data. 
As a further extension, we use not only the perturbative coupling $\alpha_s(\mu)$ 
in the evolution equations but employ also a frequently suggested ``non-perturbative'' strong coupling that approaches a constant $\alpha^{\rm{eff}}_{s,\rm{max}}$ in the infrared region.  

The second purpose of this work is to reexamine the model calculations of \cite{Myhrer:2007cf,Thomas:2008ga,Thomas:2009gk,Thomas:2008bd} and also to study possible improvements (Section \ref{modelpionOGE}). Given these results and the evolved lattice data, we conclude with a discussion in Section \ref{Discussion}.

\section{QCD evolution of lattice results}
\label{sec_evo}
In this section our aim is to evolve results from lattice QCD, usually provided in the $\MSbar$ scheme at a scale $\mu^2\simeq4\GeV^2$, down to the low scales characteristic of model calculations. 
The lattice calculations were performed on the basis of manifestly gauge invariant operators. The computations correspond to the spin decomposition proposed by Ji, Eq.~(\ref{spin operators2}).
For the remainder of this section, we will therefore employ the gauge invariant definitions of the spin observables.
We drop the superscript $\text{GI}$ in the following for better readability.
To obtain the complete set of evolution equations for all individual parts of the spin sum rule, we define the orbital angular momentum of quarks as $L_q=J_q-\frac{1}{2}\Delta \Sigma$, of gluons as $L_g=J_g-\Delta G$ 
(for discussions of the latter definition, see Refs.\cite{Ji:1996ek,Burkardt:2008jw, Wakamatsu:2010qj}).

Note that the gauge invariant $\Delta G$ cannot be represented in terms of a local operator \cite{Jaffe:1995an}, 
but it can be defined as the lowest $x$-moment of the gauge invariant gluon spin distribution, $\Delta g(x)$. Despite remarkable experimental and theoretical efforts with respect to polarized parton distributions \cite{Blumlein:2010rn,:2009ey,:2010um,Abelev:2007vt,deFlorian:2009vb,Leader:2010rb}, little is known so far about the magnitude of $\Delta G$. Concerning the numerical evaluation of the evolution equations, we will therefore concentrate on the quark spin, the quark orbital angular momentum and the total angular momentum of the gluons. As will be shown below, this can be done without explicit knowledge about $\Delta G$ and $L_g=J_g-\Delta G$. It then follows that the evolution of all quantities of interest can also be performed at NNLO, employing known results for the relevant anomalous dimensions from the literature.

The total angular momentum contributions $J_q$ and $J_g$ are introduced as in \cite{Ji:1996ek} in the framework of generalized parton distributions. 
We observe that $J_q$ and $J_g$ mix in exactly the same way under renormalization as the 
(symmetric and traceless) quark and gluon energy momentum tensors. 
This can be seen for example by rewriting
\begin{eqnarray}
\label{mastereq}
\langle P, s| J_{q,g}^i|P,s\rangle&=&\nonumber\\
&&\hspace{-25mm}\frac{1}{2}\epsilon^{ijk}\lim_{\Delta^\mu \rightarrow0}\Big[i\frac{\partial}{\partial \Delta^j} \langle P+\frac{\Delta}{2},s|T_{q,g}^{0k}|P-\frac{\Delta}{2},s\rangle\nonumber\\&&\hspace{-25mm}-i\frac{\partial}{\partial \Delta^k} \langle P+\frac{\Delta}{2},s|T_{q,g}^{0j}|P-\frac{\Delta}{2},s\rangle\Big](2\pi)^3 \delta(\vec \Delta),
\end{eqnarray}
where $P$ is the quark momentum, $s$ the quark helicity, and $\Delta$ is a momentum difference between incoming and outgoing quark.
Here, the additional derivative with respect to the momentum transfer, $\Delta^\mu$, cannot have any influence on the singular behavior of the operators. Therefore they mix in the same manner. 
The QCD evolution equations for $J_q$ and $J_g$ are constructed using the spin-2 singlet anomalous dimension given at next-to-leading order in \cite{Buras:1979yt,Floratos:1978ny} and at next-to-next-to-leading order in \cite{Larin:1993vu,Larin:1996wd,Retey:2000nq}. This yields 
\begin{eqnarray}
\frac{d}{d\ln  \mu^2} \left( \begin{array}{c} J_q\vspace{2mm}\\ J_g\end{array} \right)&=&- \frac{\alpha_s}{4\pi} \left(\begin{array}{cc} \frac{32}{9} \hspace{+2mm}& \quad-\frac{2}{3}n_F\vspace{+2mm}\\  -\frac{32}{9}  \hspace{+2mm}&\quad \frac{2}{3}n_F \end{array} \right) \left( \begin{array}{c} J_q\vspace{2mm}\\ J_g\end{array} \right)
\nonumber\\\nonumber\\
&&\hspace{-26mm}-\Big( \frac{\alpha_s}{4\pi}\Big)^2 \left(\begin{array}{cc}a_1-b_1n_F \hspace{+2mm}& \quad -d_1n_F\vspace{+2mm}\\ -a_1+b_1n_F \hspace{+2mm}&\quad d_1n_F \end{array} \right) \left( \begin{array}{c} J_q\vspace{2mm}\\ J_g\end{array} \right)\nonumber\\
&&\hspace{-26mm}-\Big( \frac{\alpha_s}{4\pi}\Big)^3\left(
\begin{array}{cc}
 a_2-b_2 n_F-c_2 n_F^2 \hspace{+2mm}& \quad-d_2 n_F+e_2 n_F^2\vspace{+2mm}\\
 -a_2+b_2n_F+c_2 n_F^2\hspace{+2mm}& \quad d_2n_F-e_2n_F^2\end{array}  \right)\left( \begin{array}{c} J_q\vspace{2mm}\\ J_g\end{array} \right)\nonumber\\
\label{JqJg}
\end{eqnarray}
\begin{table*}
\caption{\label{Konstanten}Coefficients entering the evolution equations (\ref{JqJg}).}
\centering
\begin{tabular}{cccccccc}
 $a_1$&$b_1$&$d_1$&$\quad a_2\quad$&$\quad b_2 \quad$&$\quad c_2\quad$&$\quad d_2\quad$&$\quad e_2\quad$\\ \hline
 $\frac{11744}{243}$&$\frac{416}{81}$&$\frac{611}{81}$&\vphantom{\Big(}$\frac{5514208}{6561}+\frac{1280}{81}\zeta(3)$&$\frac{134888}{2187}+\frac{2560}{27}\zeta(3)$&$\frac{1136}{243}$&$\frac{670871}{4374}-\frac{2600}{27}\zeta(3)$&$\frac{8830}{729}$\\
\end{tabular}
\end{table*}
for $n_F$ flavours (compare also \cite{Wakamatsu:2007ar}), with entries $a_i,b_i,...$ given in Table \ref{Konstanten}. 
For the non-singlet combination $J^{NS}_q$, we find Eq. (\ref{JNSx}).
\begin{figure*}
\begin{eqnarray}
\label{JNSx}
\frac{d}{d\ln  \mu^2}J_q^{NS}&=&- \frac{\alpha_s}{4\pi}\frac{32}{9}J_q^{NS}-\Big( \frac{\alpha_s}{4\pi}\Big)^2\Big(\frac{11744}{243}-\frac{256}{81} n_F\Big)J_q^{NS}\nonumber\\
&&-\Big( \frac{\alpha_s}{4\pi}\Big)^3\left(\frac{5514208}{6561}+\frac{1280\, \zeta(3)}{81}-\frac{167200 \,n_F}{2187}-\frac{1280\,n_F \,\zeta(3)}{27} -\frac{896 \,n_F^2}{729}\right)J_q^{NS}.
\end{eqnarray}
\end{figure*}
The evolution equations for the spin contributions at NNLO in the $\mbox{MS}(\overline{\mbox{MS}})$ scheme \cite{'tHooft:1973mm,Bardeen:1978yd}
(these two schemes are simply connected through a change in the renormalization scale)
are given by \cite{Mertig:1995ny,Vogt:2008yw,Vogelsang:1995vh}
 \begin{eqnarray}
 \label{eveqcomplete}
\frac{d}{d\ln \mu^2}\left(\begin{array}{c}\Delta \Sigma \vspace{2mm}\\ \Delta G\end{array}\right)&=&- \frac{\alpha_s}{4 \pi}  \left(\begin{array}{cc} 0  & \quad0 \vspace{2mm}\\ -4 & \quad-\beta_0   \end{array}\right)\left(\begin{array}{c}\Delta \Sigma\vspace{2mm}\\ \Delta G\end{array}\right)
\nonumber\\
&&\hspace{-10mm}- \Big(\frac{\alpha_s}{4 \pi}\Big)^2\left(\begin{array}{cc}8 n_F &\quad 0 \vspace{2mm} \\ -\frac{236}{3}+\frac{8}{9} n_F & \quad-\beta_1  \end{array}\right) \left(\begin{array}{c}\Delta \Sigma \vspace{2mm}\\ \Delta G\end{array}
\right)\nonumber\\
&&\hspace{-10mm}- \Big(\frac{\alpha_s}{4 \pi}\Big)^3\left(
\begin{array}{cc}
\frac{472}{3}n_F-\frac{16 n_F^2}{9} & \quad0 \vspace{1mm}\\
 \gamma_{gq} &\quad \gamma_{gg}
\end{array}
\right)\left(\begin{array}{c}\Delta \Sigma \vspace{2mm}\\ \Delta G\end{array}\right)\,.\nonumber\\
\end{eqnarray}
At NNLO, the anomalous dimensions $\gamma_{gq}$ and $\gamma_{gg}$ are still unknown, while the upper row ($\gamma_{qq},\gamma_{qg}$) has been obtained as described in \cite{Vogt:2008yw}.
Here the QCD beta functions are
\begin{equation}
\beta_0=11-\frac{2n_F}{3} ,\quad \quad \beta_1=102-\frac{38}{3} n_F.
\end{equation}
We emphasize that in the chosen renormalization scheme, the evolution of $\Delta\Sigma$ is independent of $\Delta G$, even at NNLO. 
Furthermore, since $J_q+J_g=1/2$ at any scale, the evolution of $J_q$ does not require an independent knowledge of the value of $J_g$.
Hence one finds the remarkable result, already mentioned above, that neither $\Delta G$ nor $L_g=J_g-\Delta G$ 
are actually required in practice for the scale evolution of $L_q=J_q-\Delta\Sigma/2$. As a consequence the evolution of all the quantities in Eq.~(\ref{spin operators2}) can be performed at NNLO.

Employing the definitions of $L_q$ and $L_g$ given above, fully consistent coupled evolution equations for the orbital angular momenta of quarks and of gluons can be written,
\begin{eqnarray}
 \label{eveqcomplete2}
 \frac{d}{d\ln \mu^2}\left(\begin{array}{c}L_q \vspace{+2mm}\\ L_g \end{array}\right)&=&- \frac{\alpha_s}{4 \pi}  \left(\begin{array}{cc} \frac{32}{9} &\quad -\frac{2}{3}n_F  \vspace{+2mm}\\ -\frac{32}{9} & \quad\frac{2}{3}n_F   \end{array}\right)\left(\begin{array}{c}L_q \vspace{+2mm}\\ L_g\end{array}\right)
\nonumber\\&&- \frac{\alpha_s}{4 \pi}  \left(\begin{array}{cc}  \frac{16}{9} &\quad -\frac{2}{3}n_F  \vspace{+2mm}\\ \frac{20}{9} &\quad\mbox{\footnotesize{11}}    \end{array}\right)
 \left(\begin{array}{c}\Delta \Sigma \vspace{+2mm}\\ \Delta G\end{array}\right)\nonumber\\
&&\hspace{-20mm}- \Big(\frac{\alpha_s}{4 \pi}\Big)^2  \left(\begin{array}{cc} \frac{11744}{243}-\frac{416}{81}n_F &\quad -\frac{611}{81}n_F  \vspace{+2mm}\\ -\frac{11744}{243}+\frac{416}{81}n_F &\quad \frac{611}{81}n_F   \end{array}\right)\left(\begin{array}{c}L_q \vspace{+2mm}\\ L_g\end{array}\right)\vspace{+3mm}\nonumber\\\nonumber\\
&&\hspace{-20mm}- \Big(\frac{\alpha_s}{4 \pi}\Big)^2  \left(\begin{array}{cc} \frac{5872}{243}-\frac{532}{81}n_F &\quad -\frac{611}{81}n_F  \vspace{+2mm}\\ \frac{13244}{243}+\frac{136}{81}n_F & \quad\mbox{\footnotesize{102}}-\frac{415}{81}n_F     \end{array}\right)\left(\begin{array}{c}\Delta \Sigma \vspace{+2mm}\\ \Delta G\end{array}\right),\nonumber\\
 \end{eqnarray}
at next-to-leading order in the $\mbox{MS}(\overline{\mbox{MS}})$ scheme.

An overview of lattice QCD calculations of nucleon spin observables, in particular of moments of generalized
parton distributions that give access to the total quark angular momentum $J_q$, can be found in \cite{Hagler:2009ni}.
Here we focus on the latest published results from the LHP collaboration \cite{Bratt:2010jn}.
They were obtained in the framework of a mixed action approach with $n_F=2+1$ dynamical fermions, with 
lattice pion masses as low as $\approx300\MeV$. 
The computationally demanding quark line disconnected diagrams, 
which contribute in the singlet sector, were not included in this study. 
The final values for $\Delta\Sigma_q$, $L_q$ and $J_q$ at the physical pion mass were obtained from extrapolations
employing the covariant baryon chiral perturbation theory results of \cite{Dorati:2007bk}.
We refer to the original publication \cite{Bratt:2010jn} for the details of the lattice simulation, the numerical analysis,
and a discussion of the statistical and systematic uncertainties. A summary of the lattice results, for the $\MSbar$ scheme at a scale of $4\GeV^2$, is given in Table \ref{Lattice}. The errors given in this table do not include systematic uncertainties from chiral extrapolations and disconnected diagrams. 

\begin{figure*}
\hspace{0mm}
\begin{centering}
\begin{minipage}{0.49\linewidth}
\footnotesize{a.)$\qquad\qquad\qquad\qquad\qquad\qquad\qquad\qquad$}
\vspace{-3mm}
\begin{center}
\end{center}
\includegraphics[width=0.88\linewidth]{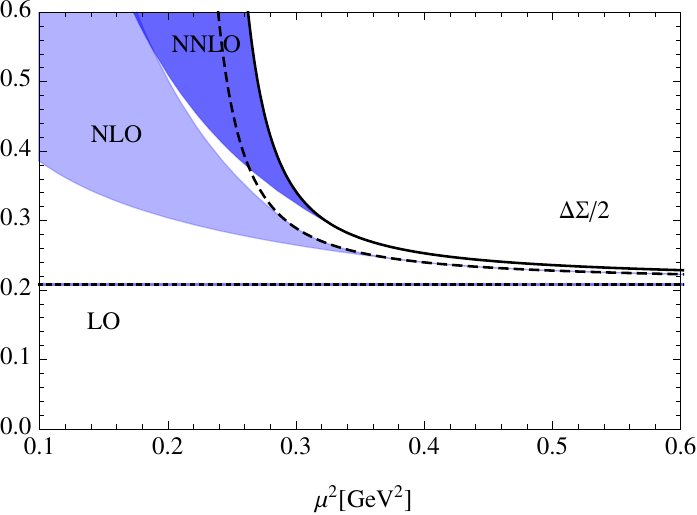}%
\end{minipage}
\begin{minipage}{0.49\linewidth}
\footnotesize{b.)$\qquad\qquad\qquad\qquad\qquad\qquad\qquad\qquad$}
\vspace{-3mm}
\begin{center}
\end{center}
\hspace{-3mm}\includegraphics[width=0.9\linewidth]{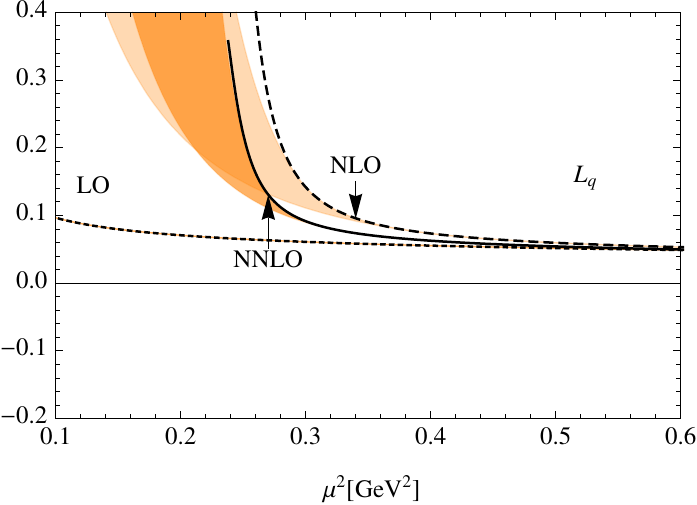}
\end{minipage}
\begin{minipage}{0.49\linewidth}
\footnotesize{c.)$\qquad\qquad\qquad\qquad\qquad\qquad\qquad\qquad$}
\vspace{-3mm}
\begin{center}
\end{center}
\includegraphics[width=0.9\linewidth]{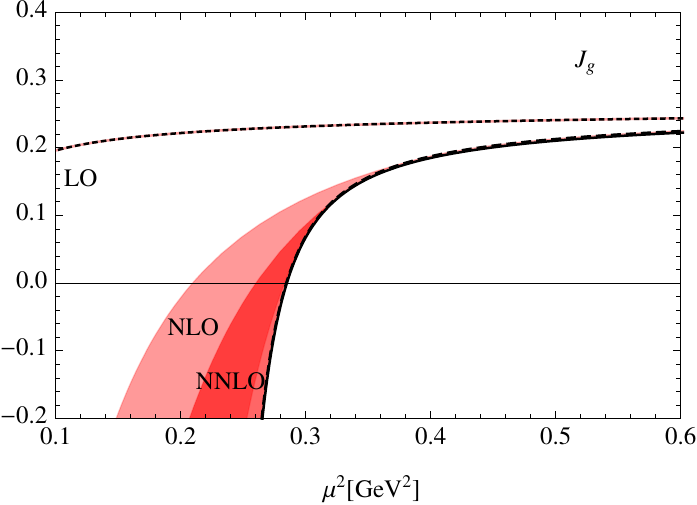}%
\end{minipage}
\begin{minipage}{0.49\linewidth}
\footnotesize{d.)$\qquad\qquad\qquad\qquad\qquad\qquad\qquad\qquad$}
\vspace{-3mm}
\begin{center}
\end{center}
\includegraphics[width=0.9\linewidth]{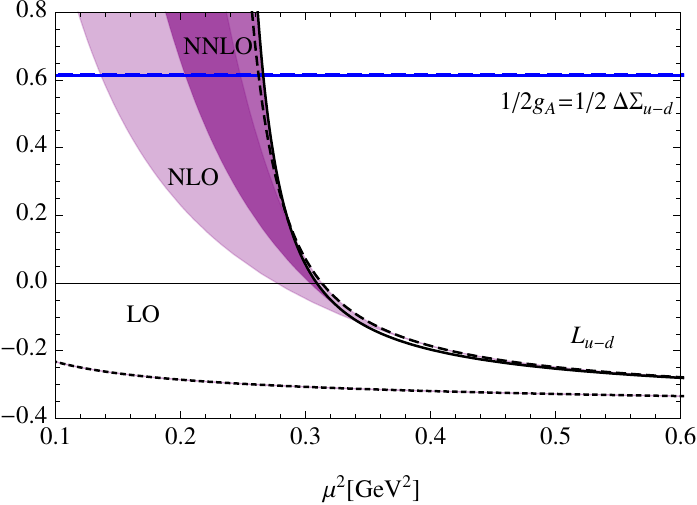}
\end{minipage}
\end{centering}
\caption{\label{evolutionall}Scale dependence of $\Delta \Sigma/2$, $L_q$, $J_g$, and $L_{u-d}$ shown together with $\frac{1}{2} g_A=\frac{1}{2}\Delta \Sigma_{u-d}$, starting from the lattice QCD results at $\mu^2=4\,\rm{GeV}^2$ given in Table \ref{Lattice}. In all diagrams the solid, dashed and short-dashed black lines are solutions of the QCD evolution equations at NNLO, NLO, and LO, respectively. 
The colored bands are obtained by imposing upper bounds for $\alpha_s$ (see text).
}
\end{figure*}

\begin{table}
\caption{\label{Lattice}Lattice QCD results from Ref.~\cite{Bratt:2010jn} for $\Delta \Sigma/2$ and $L_q$ 
in the $\MSbar$ scheme at $\mu^2=4\GeV^2$, separated into u- and d-quark contributions. Statistical and estimated systematic uncertainties due to the renormalization are given in the form $(\ldots)_{\rm{stat}}$$(\ldots)_{\rm{ren}}$.}
\centering
\begin{tabular}{c|cc}
\vphantom{\Big(}&$\Delta \Sigma/2$&$L_q$\\
\hline
\vphantom{\Big(}$\qquad$u$\qquad$&0.411(36)& -0.175(36)(17)\\
\vphantom{\Big(}$\qquad$d$\qquad$&-0.203(35)& 0.205(35)(0)\\
\end {tabular}
\end{table}

\begin{figure}[!t]
\includegraphics[width=0.92\linewidth]{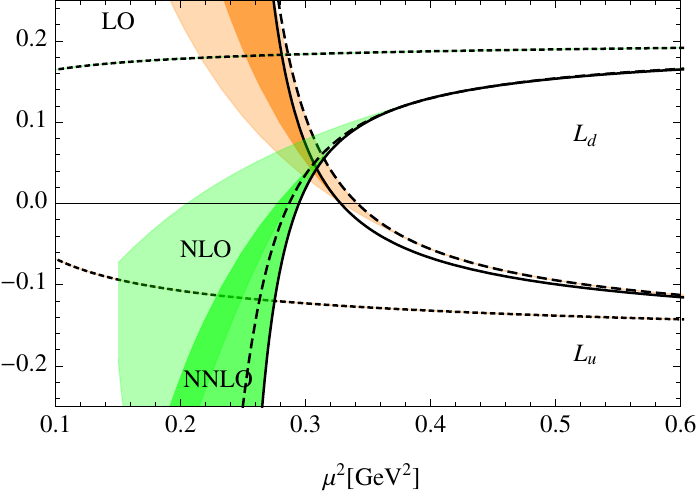}%
\caption{\label{evolutionLuLd}Evolution of $L_u$ and $L_d$ in NNLO (solid lines), NLO (dashed lines) and LO (short dashed lines). The 
shaded areas are assigned as in Fig.~\ref{evolutionall}.}
\end{figure}

\begin{figure*}
\hspace{0mm}
\begin{centering}
\begin{minipage}{0.49\linewidth}
\footnotesize{a.)$\qquad\qquad\qquad\qquad\qquad\qquad\qquad\qquad$}
\vspace{-3mm}
\begin{center}
\end{center}
\includegraphics[width=0.88\linewidth]{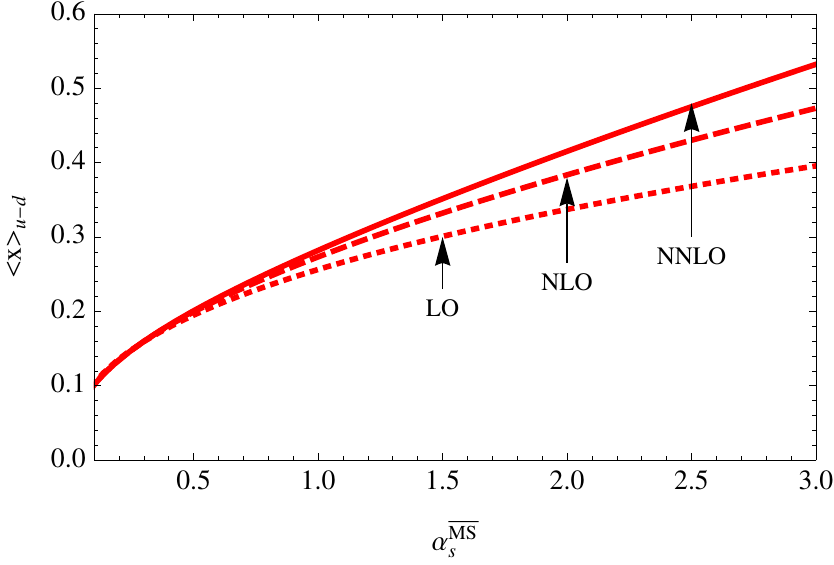}%
\end{minipage}
\begin{minipage}{0.49\linewidth}
\footnotesize{b.)$\qquad\qquad\qquad\qquad\qquad\qquad\qquad\qquad$}
\vspace{-3mm}
\begin{center}
\end{center}
\hspace{-3mm}\includegraphics[width=0.9\linewidth]{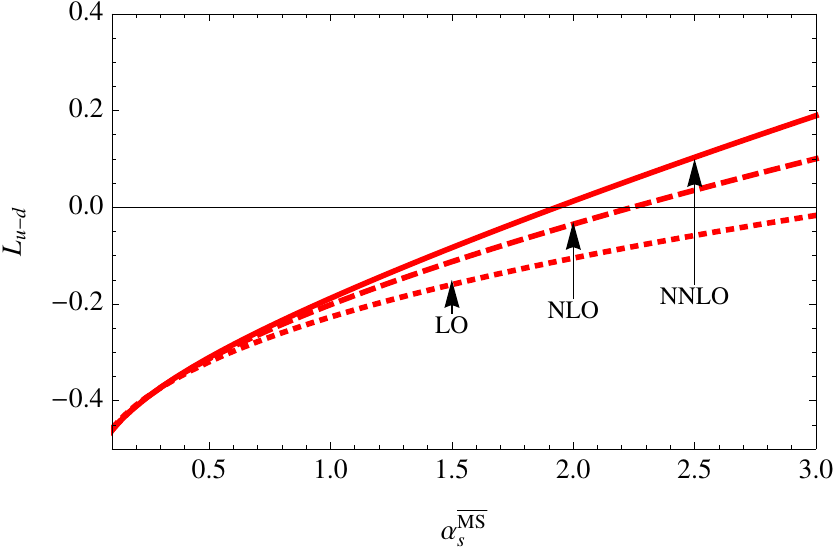}
\end{minipage}
\begin{minipage}{0.49\linewidth}
\footnotesize{c.)$\qquad\qquad\qquad\qquad\qquad\qquad\qquad\qquad$}
\vspace{-3mm}
\begin{center}
\end{center}
\includegraphics[width=0.9\linewidth]{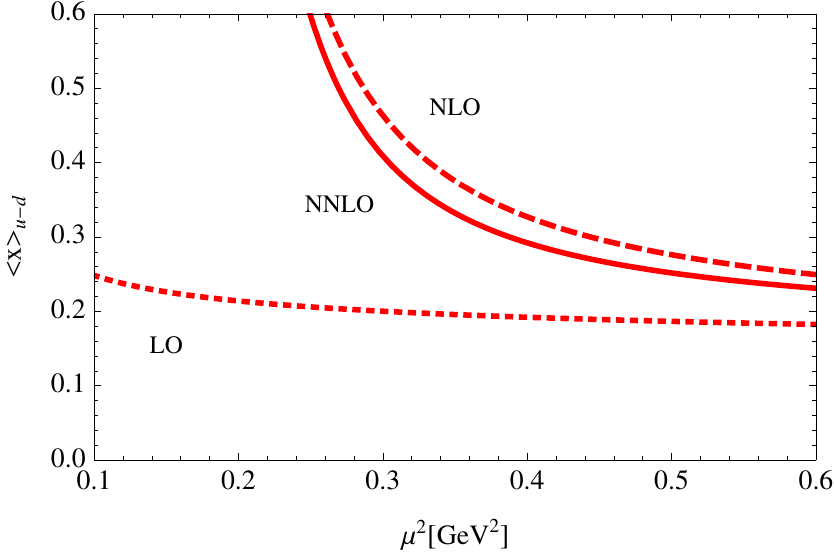}%
\end{minipage}
\begin{minipage}{0.49\linewidth}
\footnotesize{d.)$\qquad\qquad\qquad\qquad\qquad\qquad\qquad\qquad$}
\vspace{-3mm}
\begin{center}
\end{center}
\includegraphics[width=0.9\linewidth]{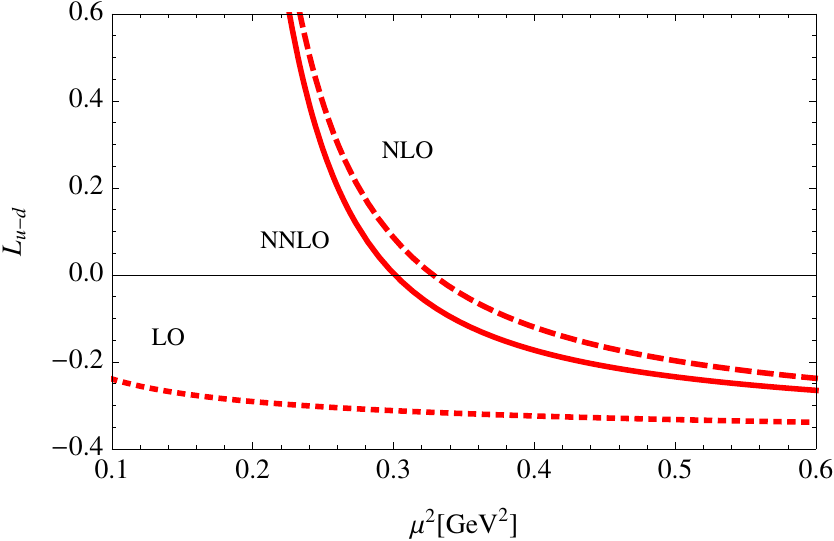}
\end{minipage}
\end{centering}
\caption{\label{alphadep}$\alpha_s$-dependence of the momentum and angular momentum contributions $\langle x\rangle_{u-d}$ and $L_{u-d}$ with starting values fixed at $\alpha_{s,0}=0.28$. Diagrams c and d show the corresponding renormalization scale dependence. In all the diagrams solid, dashed and short-dashed red lines correspond to  NNLO, NLO, and LO, respectively.}
\end{figure*}

For our QCD evolution of lattice results, we assume a vanishing contribution from strange quarks (see also \cite{Bass:2009ed}). The total gluon angular momentum is given by $J_g=\frac{1}{2}-J_{u+d}$. 
As starting values at $\mu^2=4\, \rm{GeV}^2$, we use the numbers given in Table \ref{Lattice}. For the running coupling, we set $\alpha_s^{\rm{NNLO},n_\textsl{F}=5}(M_Z)=0.1184$ \cite{Nakamura:2010zzi}, and employ the flavor matching conditions \cite{Bethke:2009jm,Chetyrkin:1997sg} to obtain $\Lambda^{\rm{NNLO,n_\textsl{F}=3}}_{\rm{QCD}}=0.338\,\rm{GeV}$, $\Lambda^{\rm{NLO,n_\textsl{F}=3}}_{\rm{QCD}}=0.388\,\rm{GeV}$, and $\Lambda^{\rm{LO,n_\textsl{F}=3}}_{\rm{QCD}}=0.148$ $\rm{GeV}$. With this input, we have solved the LO, NLO and NNLO coupled evolution equations and found the scale dependence plotted in Figures \ref{evolutionall} and \ref{evolutionLuLd}. 

The results at LO, NLO and NNLO, employing the standard analytical expressions for the perturbative strong coupling constant
(corresponding to an expansion in \linebreak$1/\ln(\mu^2/\Lambda^2)$ beyond LO, see, e.g., \cite{Nakamura:2010zzi}) in the $\MSbar$ scheme 
at the appropriate order, are given by the short-dashed, dashed, and solid 
 black curves, respectively.
We note that the deviation of the approximate analytical expressions for $\alpha_s$ from the exact (numerical) solutions
of the evolution equations increases as one approaches lower scales. 
For $n_F=3$ and $\Lambda^{\rm{NNLO}, n_\textsl{F}=3}_{\rm{QCD}}=0.338\,\rm{GeV}$,
the formally exact solution for the running coupling at NNLO would already diverge around $\mu^2\sim 0.27\GeV^2$. 
The curves in Fig.~\ref{evolutionall} obtained for $\alpha_s$ in the $1/\ln(\mu^2/\Lambda^2)$-approximation are therefore only indicative 
for a strong coupling constant that grows indefinitely as $\mu^2\rightarrow 0$.

A comparison with the model results, e.g. as proposed by Myhrer and Thomas \cite{Myhrer:2007cf,Thomas:2008ga,Thomas:2009gk,Thomas:2008bd}, requires an evolution down to scales $\mu^2 \sim 0.1 - 0.3\, \rm{GeV}^2$, far away from the perturbative QCD regime. 
From the results in Figs. \ref{evolutionall} and \ref{evolutionLuLd}, we find that the evolution curves at
NLO and NNLO begin to show a very strong curvature exactly in this region. Clearly, at such low scales quantitative statements based 
on a perturbative QCD analysis (including the running of $\alpha_s$) are 
no longer reliable.

With respect to (the non-perturbative) $\alpha_s$, one would expect in any case that it saturates at low scales,
as suggested by non-perturbative resummation in the infrared region \cite{Shirkov:1997wi, Fischer:2003rp, Prosperi:2006hx}.
A further rough impression about the uncertainties in the evolution may therefore be
obtained as follows. As an alternative to the infrared divergent, perturbative
coupling  $\alpha_s(\mu^2)$ in the evolution equations, we consider an effective  $\alpha^{\rm{eff}}_s(\mu)$ 
that approaches a fixed value $\alpha^{\rm{eff}}_{s,\rm{max}}$ at small $\mu^2$. 
For the corresponding numerical calculation we use $\alpha_{s}^{\rm{eff}}(\mu^2)=\alpha_s(\mu^2)$ of appropriate order in the $\MSbar$ scheme for all $\mu$ for which $\alpha_{s}(\mu^2)\leqslant \alpha^{\rm{eff}}_{s,\rm{max}}$. Below the scale $\mu_0$ at which $\alpha_{s}(\mu_0^2)=\alpha_{s,\rm{max}}^{\rm{eff}}$, we use $\alpha^{\rm{eff}}_{s}\equiv\alpha_{s,\rm{max}}^{\rm{eff}}$. 
For illustration, we chose two different values, $\alpha^{\rm{eff}}_{s,\rm{max}}=1.5$ and $\alpha^{\rm{eff}}_{\rm{s,max}}=3$. Performing the evolution with these restricted couplings spans the shaded colored bands in Figures \ref{evolutionall} and \ref{evolutionLuLd}. The boundary with flatter evolution always corresponds to $\alpha^{\rm{eff}}_{s,\rm{max}}=1.5$, the steeper one to $\alpha^{\rm{eff}}_{s,\rm{max}}=3$. The lighter colored bands correspond to NLO, the darker colored bands to NNLO. 
The relatively small $\Lambda^{\rm{LO,\,n_F=3}}_{\rm{QCD}}=0.148\,\rm{GeV}$, obtained from the flavor matching procedure, implies that the corresponding
$\alpha_{s}(\mu^2)$ at LO stays below $\alpha_{s,\rm{max}}^{\rm{eff}}$ in the considered region of $\mu^2$. Hence 
no bands are shown in this case, and one finds the remarkably stable (but unrealistic) LO evolution shown in the figures.

NLO and NNLO evolution results in the dashed and solid lines, and the lighter and darker shaded colored bands, respectively. 
As mentioned before, in contrast to the LO evolution, strong evolution effects can already be seen at scales $\mu^2\lesssim0.3\,\rm{GeV}^2$.  
Before discussing the results in more detail, we note as a general feature 
that the bands obtained for our choices of $\alpha_{s,\rm{max}}^{\rm{eff}}$ start to broaden quickly below
$\mu^2\sim0.3\,\rm{GeV}^2$, indicating potentially large uncertainties in the evolution as one enters the non-perturbative regime.
With the exception of $\Delta\Sigma$, we observe a broad overlap of the NLO and NNLO bands 
for each of the different observables.

At these orders we find interesting qualitative and quantitative changes 
of the proton spin decomposition under evolution.
The singlet quark spin contribution $\Delta \Sigma$ 
becomes scale dependent at NLO and increases with lower scales, a behavior that is even more strongly pronounced in NNLO.
Similarly, the contribution from $L_q$ stays positive and grows quickly at low scales, while
$J_g$ crosses zero in the region of $\mu^2\sim0.21$ to $0.29\,\rm{GeV}^2$ and then becomes large and negative.
At the same time, $L_{u-d}$ \footnote{showing less systematic uncertainties in the lattice computation as contributions from disconnected 
diagrams cancel out for isovector quantities} 
shows a strong upwards bending and moves from its negative starting value at higher scales towards
large positive values at low scales, crossing zero around $\mu^2\sim0.28$ to $0.32\,\rm{GeV}^2$.
From the evolution of $L_q$ and $L_{u-d}$ one can deduce the separate $\mu$-dependences of $L_u$ and $L_d$, as displayed in Figure~\ref{evolutionLuLd}. Both $L_u$ and $L_d$ change sign under evolution.
As one moves in the direction of lower scales, the contribution from the up-quarks, $L_u$, changes from negative to increasingly large positive values
at about $\mu^2\sim0.32-0.35\,\rm{GeV}^2$, while the zero crossing of $L_d$ from positive to increasingly large negative values takes places at
slightly lower scales of $\mu^2\sim0.2$ to $0.3\,\rm{GeV}^2$.
Similar trends have already been observed in the LO-study of Ref.~\cite{Thomas:2008ga}, based however on a substantially larger $\Lambda^{\rm{LO},n_\textsl{F}=3}_{\rm{QCD}}=0.24\,\rm{GeV}$. 
It is interesting to observe
that the crossing points of $L_u$ and $L_d$ roughly match
as we proceed from 
NLO to NNLO. 

For a better understanding of the scale dependence at very low scales, i.e. for large values of the strong coupling constant,
it is instructive to present the evolution 
in terms of $\alpha_s$ instead of $\mu$.\footnote{Ph.H. would like to thank
M. Diehl for helpful discussions on this point.} 
This is illustrated in Figure \ref{alphadep} for the nucleon isovector momentum fraction $\langle x\rangle_{u-d}$ on the left and for $L_{u-d}$ on the right. 
Note that the evolution equations for both quantities are based on the same anomalous dimension. 
The results were obtained by rewriting the evolution equations,
i.e.
translating derivatives with respect to $\mu^2$ 
into derivatives with respect to $\alpha_s$, and then replacing $\frac{d\alpha_s}{d\ln\mu^2}$ 
by the QCD $\beta$-function. 
As starting values we have used $\langle x\rangle_{u-d}=0.155$ and $L_{u-d}=-0.38$ \cite{Bratt:2010jn} at $\alpha_{s,0}=0.28$ (corresponding to a scale of $\mu^2\sim4\rm{GeV}^2$). 
Both quantities are remarkably stable under evolution in $\alpha_s$, even at large $\alpha_s$. 
For example, at $\alpha_s=2.0$ the momentum fraction $\langle x\rangle_{u-d}$ at NLO is about $\sim14\%$ larger than the LO result, and
NNLO and NLO results differ by only about $\sim8\%$. 
Apart from the difference in the starting values, the form of the $\alpha_s$-dependence is identical
for $L_{u-d}$ and $\langle x\rangle_{u-d}$. The zero crossing of $L_{u-d}$ quickly moves to smaller $\alpha_s$ as the order of the perturbative evolution is increased. 
Figures \ref{alphadep}c and d show how the $\alpha_s$ dependences translate to a renormalization scale dependence. 
The logarithmic dependence of $\alpha_s(\mu)$ that shows up strongly as $\mu$ approaches $\Lambda_{\rm{QCD}}$,
produces the strong curvature in the evolution of the observables as functions of the scale $\mu$.

Small but unimportant differences between Fig. 3d and Fig. 1d result from the slightly different procedures involved, as described.

Further remarks on the evolution
and a comparison of evolved lattice results with calculations performed in a chiral quark model will be presented below in Section \ref{Discussion}.

\section{Contributions to the nucleon spin in a chiral quark model}
\label{modelpionOGE}
\subsection{Pion cloud contributions, revisited}

\label{pions}
It is a well established fact that the nucleon is a complex many-body system, with the three valence quarks and multiple quark-antiquark pairs embedded in a strong, non-perturbative gluonic field configuration. Chiral quark models draw a simplified picture of this complexity in terms of valence quarks in a confining bag coupled to the pion cloud, based on spontaneously broken chiral symmetry in low-energy QCD. A frequently used representative of such chiral models is the cloudy bag (CBM) 
\cite{Thomas:2001kw,Thomas:1982kv,Theberge:1982xs} that couples the pion cloud to quarks in the MIT bag \cite{Chodos:1974pn} such that chiral invariance is realized in the limit of massless quarks. This section summarizes the present status concerning nucleon spin structure from this model point of view.

The relativistic treatment of quarks itself yields already results that differ significantly from the ordinary $\rm{SU(6)}$ quark model predictions. 
The $\Delta \Sigma =1$ of the non-relativistic quark model is reduced to about $\Delta \Sigma^{\rm{MIT}}=0.65$ in the MIT bag model. The ``missing spin'' is interpreted as orbital angular momentum of the valence quarks, $2L^{\rm{MIT}}_{u+d}=0.35$, associated with the lower components of the Dirac quark wave functions.

\begin{figure}
\centering
\includegraphics[width=0.8\linewidth]{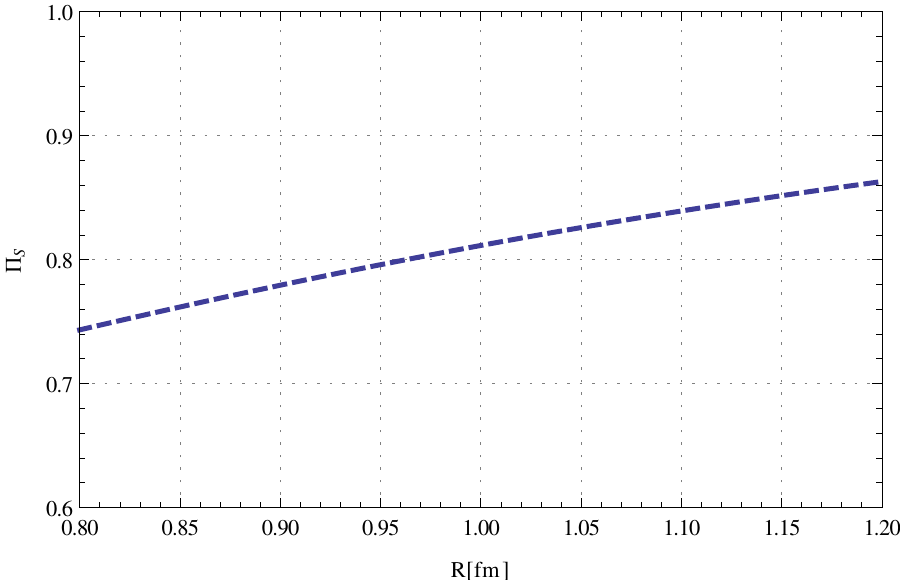}
\includegraphics[width=0.8\linewidth]{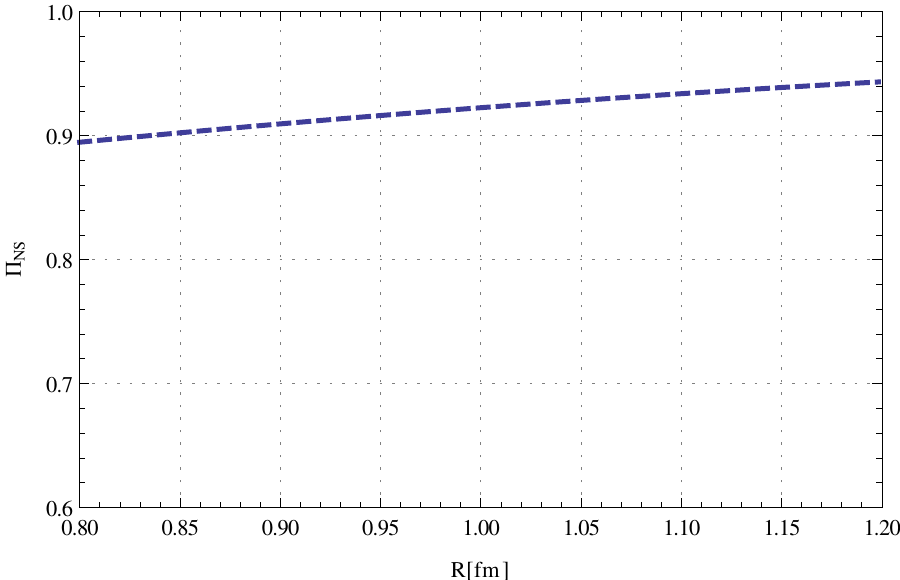}
\caption{Radius dependence of the singlet and non-singlet corrections $\Pi_{\rm{S}}$ and $\Pi_{\rm{NS}}$ associated with the pion cloud of the nucleon.}
\label{PiS}
\end{figure}

The correction factors for the pion cloud in the CBM were already derived by Myhrer and Thomas in \cite{Myhrer:2007cf,Thomas:2008ga,Thomas:2009gk,Thomas:2008bd}. For the singlet expectation values, 
\begin{eqnarray}
\label{singlet}
\Delta \Sigma^{\rm{CBM}}&=&0.65\cdot \Pi_{\rm{S}}(R),\qquad 2 L_q^{\rm{CBM}}=0.35 \cdot \Pi_{\rm{S}}(R),\nonumber\\
&&\qquad 2L_\pi^{\rm{CBM}}=1-\Pi_{\rm{S}}(R),
\end{eqnarray}
 and for the non-singlet expectation values \cite{Bass:2009ed},
 \begin{eqnarray}
 \label{non-singlet}
\Delta \Sigma^{\rm{CBM}}_{u-d}&=&g^{(3)\;\rm{CBM}}_A=\frac{5}{3}\cdot 0.65\cdot \Pi_{\rm{NS}}(R),\nonumber\\
2 L^{\rm{CBM}}_{u-d}&=&\frac{5}{3}\cdot 0.35\cdot \Pi_{\rm{NS}}(R)\,.
\end{eqnarray}
We have denoted the pion cloud correction factors by $\Pi_{\rm{S}}(R)$ and $\Pi_{\rm{NS}}(R)$, each for a given bag radius $R$. 
For their explicit form we refer to \cite{Myhrer:2007cf,Thomas:2008ga,Thomas:2009gk,Thomas:2008bd, Bass:2009ed}  and references therein. 
We have reproduced these factors using the formalism described in \cite{Theberge:1982xs}. Their radius dependence is plotted in Figure \ref{PiS}. 
The singlet correction factor, $\Pi_{\rm{S}}$, is smaller than unity and 
hence leads to the expected reduction of the quark spin contribution.
At the same time, $\Pi_{\rm{NS}}<1$ leads to a slightly less favourable comparison of $g^{(3)\;\rm{CBM}}_A$
with the experimental value of $g_A$. This mismatch is a feature that depends on the chiral representation chosen for the model. In particular, as pointed out in \cite{Bass:2009ed}, choosing a volume coupling version instead of the standard surface coupling reproduces the experimental value of $g_A^{(3)}$ with very good accuracy.
A common way of treating the discrepancy between $g_A^{(3)\rm{CBM}}$ and the empirical $g_A\equiv g^{(3)}_A=1.27$ is by inclusion of a phenomenological center-of-mass correction. This correction is just a multiplicative factor for $\Delta \Sigma$ and $\Delta \Sigma_{u-d}$. \footnote{Notice that the value $g_A^{(3)}=1.27$, given in \cite{Myhrer:2007cf,Thomas:2008ga,Thomas:2009gk,Thomas:2008bd}, is obtained by adjusting the phenomenological center-of-mass correction, while this correction has not been included for any of the other spin observables listed in these references.} For consistency, however, one should then also rescale $L_q$ and $L_{u-d}$ accordingly to keep the spin sum rule conserved. 
The corresponding results are shown in Table \ref{model}, in addition to results obtained without the center-of-mass corrections. If explicit gluon operators are taken into account (see Table \ref{modelgauge} and section \ref{gluons} below), these corrections cannot be applied uniquely. In that case we give only the model results without rescaling by c.m. corrections.

A detailed analysis of $\Delta \Sigma$ also requires, in principle, a discussion of $g_A^{(8)}$ since the flavour singlet axial vector coupling constant extracted from polarised deep inelastic scattering is sensitive to that value. In the present case we restrict ourselves to a flavor-SU(2) cloudy bag model which implies $\Delta \Sigma=g_A^{(8)}$. This is quite compatible with the small strange quark contribution $\Delta s \sim -0.01$ discussed in Ref.~\cite{Bass:2009ed}.

\subsection{Corrections from one-gluon exchange processes}
\label{gluons}

The MIT bag model produces degenerate masses of the nucleon and the delta, whereas the empirical mass splitting is about $300\,\rm{MeV}$. In order to account for this mass difference, an additional spin-spin interaction between quarks is introduced. The traditional ad-hoc way of doing this is by allowing one-gluon exchanges between quarks in the interior of the bag, with an effective coupling $\tilde \alpha_s$. This coupling should not be confused with the strong coupling of QCD. It represents a free parameter chosen in such a way that the model reproduces the light hadron spectrum \cite{DeGrand:1975cf}. The values used in the literature vary between $\tilde\alpha_s=1$ \cite{Schumann:2000uw} and $\tilde\alpha_s=2.2$ \cite{Hogaasen:1987nj}. We use these two values as options in the results shown later in Table \ref{model} and \ref{modelgauge}.

\begin{figure}
\centering
\includegraphics[width=0.8\linewidth]{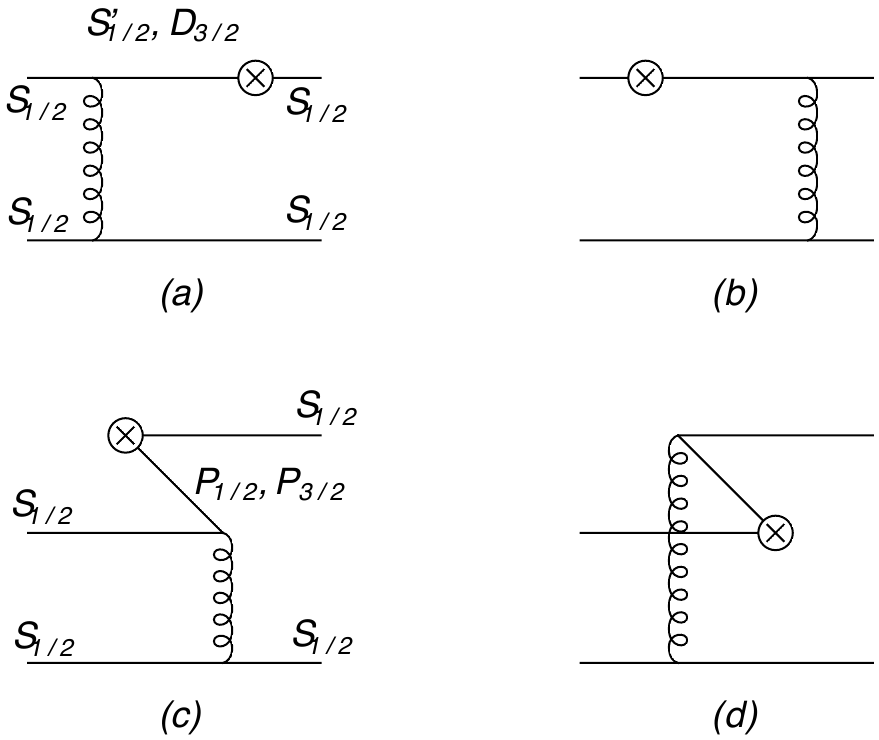}
\caption{One-gluon exchange (OGE) corrections for $\Delta \Sigma$, $L_{S,NS}$ and $g_A^{(3)}$. Diagram (a) and (b) are contribution from intermediate quark states and (c) and (d) from quark-antiquark pairs.}
\label{OGE}
\end{figure}

We have performed calculations in analogy to Ref.~\cite{Hogaasen:1987nj}, where the color magnetic corrections to baryon magnetic moments and to semileptonic decays, i.e. the axial coupling constant, were derived at order $\tilde\alpha_s$. Following the arguments given there, we neglect the color electric corrections and drop loop diagrams. 
That means, we consider diagrams \ref{OGE}(a)-\ref{OGE}(d), 
in which only color magnetic gluons are exchanged.

For the singlet expectation values, $\Delta \Sigma$ and $L_q$, we find the (additive) OGE corrections
\begin{equation}
\delta \Delta \Sigma=-2\delta_g\cdot \tilde\alpha_s,\qquad \delta L_q=\delta_g\cdot \tilde\alpha_s,
\label{resultsDez}
\end{equation}
with $\delta_g\simeq2.5\cdot 10^{-2}$, where $L_q$ is used in its non-gauge-invariant formulation (\ref{spin operators}), and intermediate (anti-)quarks in the orbitals $P_{1/2}, P_{3/2}, D_{3/2}, S'_{1/2}, P'_{1/2}, P'_{3/2}, D'_{3/2}, S''_{1/2}$ are taken into account (conventions are chosen as in \cite{Hogaasen:1987nj}). 
As already pointed out in \cite{Myhrer:2007cf,Thomas:2008ga,Thomas:2009gk,Thomas:2008bd} the corrections are mainly due to antiquarks propagating in the $P_{1/2},  P_{3/2}$ orbitals. 
Compared to $\delta \Delta \Sigma \sim -0.15$ and $\delta L_q\sim 0.08$ for $\tilde\alpha_s=2.2$ as presented in \cite{Thomas:2008ga,Thomas:2009gk}, our corrections are slightly smaller.

For the non-singlet operators we find:
\begin{equation}
\delta \Delta \Sigma_{u-d} =\delta g_A^{(3)}=\frac{2}{3} \delta_g\cdot \tilde \alpha_s,\qquad \delta L_{u-d}=-\frac{1}{3} \delta_g \cdot \tilde \alpha_s.
\label{resultsDez2}
\end{equation}
At this point our model agrees with that of Refs. \cite{Myhrer:2007cf,Thomas:2008ga,Thomas:2009gk,Thomas:2008bd}. For comparison, explicit numbers for the singlet and non-singlet contributions to the nucleon spin in the MIT bag model, as well as the OGE-improved 
MIT bag and cloudy bag model, for two different values of $\tilde\alpha_s$, are displayed in Table \ref{model}. \footnote{We tabulate these results here only for historical reasons as they are based on the non-gauge-invariant decomposition, Eq. (\ref{eq-spincontrib}).}
\begin{table*}[!t]
\caption{\label{model}Spin structure of the nucleon in the MIT bag model, with corrections from one-gluon exchanges (OGE), from the pion cloud and from center-of-mass rescaling. The non-gauge-invariant decomposition of the nucleon spin, Eqs.~(\ref{eq-spincontrib}), is used here.}
\centering
\begin{tabular}{c|cccc}
&$\Delta \Sigma/2$&$L_q$&$\Delta \Sigma_{u-d}/2$&$L_{u-d}$\\
\hline
relativistic (MIT bag model)&0.33&0.17&0.54&$0.29$\\
+OGE ($\tilde\alpha_s$=1.0): &0.30&0.20&0.55&0.28\\
 $\qquad\quad\;$($\tilde\alpha_s$=2.2):&0.27&0.23&0.56&0.27\\
+pion cloud ($R=1$fm, $\tilde\alpha_s$=1.0):&0.24&0.26&0.51&0.26\\
$\qquad\qquad\quad\;\;$($R=1$fm, $\tilde\alpha_s$=2.2):&0.22&0.28&0.52&0.25\\
\hline
+center of mass ($R=1$fm, $\tilde\alpha_s$=1.0):&0.30&0.20&0.64&0.20\\
\qquad\qquad\quad\quad\quad\;($R=1$fm, $\tilde\alpha_s$=2.2):&0.27&0.23&0.64&0.21\\
\end{tabular}
\end{table*}

\begin{centering}
\begin{table*}[!t]
\caption{Spin structure of the nucleon, based on manifestly gauge invariant operators, in the MIT bag model, 
together with corrections from one gluon exchanges (OGE) and from the pion cloud. 
}
\label{modelgauge}
\centering
\begin{tabular}{c|ccccc}
&$\Delta \Sigma/2$&$L^{\text{GI}}_q$&$J^{\text{GI}}_g$&$\Delta \Sigma_{u-d}/2$&$L^{\text{GI}}_{u-d}$\\
\hline
relativistic (MIT bag model)&0.33&0.17&0&0.54&$0.29$\\
+OGE ($\tilde\alpha_s$=1.0): &0.30&0.40&-0.20&0.55&0.21\\
 $\qquad\quad\;$($\tilde\alpha_s$=2.2):&0.27&0.68&-0.45&0.56&0.12\\
+pion cloud ($R=1$fm, $\tilde\alpha_s$=1.0):&0.24&0.42&-0.16&0.51&0.19\\
$\qquad\qquad\quad\;\;$($R=1$fm, $\tilde\alpha_s$=2.2):&0.22&0.64&-0.36&0.52&0.10\\
\end {tabular}
\end{table*}
\end{centering}

\begin{figure}
\centering
\includegraphics[width=0.4\linewidth]{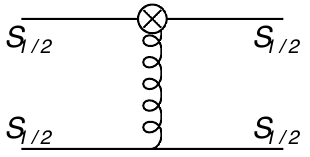}
\caption{One-gluon exchange diagram attached to a quark-quark-gluon interaction vertex.}
\label{OGEgauge}
\end{figure}

Once explicit gluon degrees of freedom, e.g. in form of one gluon exchange processes, are taken into account, 
questions of gauge invariance of the calculation must be carefully addressed.
For a consistent calculation that includes gluon exchange contributions at order $\tilde \alpha_s$,
and for a meaningful comparison with results from lattice QCD, we have to employ the gauge-invariant
orbital angular momentum operator $L^{\text{GI}}_{q}$ in Eq.~(\ref{spin operators2}) instead of 
$L_q$ defined in (\ref{spin operators}).
The covariant derivative in $L^{\text{GI}}_{q}$ produces an additional quark-gluon interaction so that one must take into account the diagram in Fig.~\ref{OGEgauge} for the corrections at order $\tilde\alpha_s$ to the MIT bag expectation values. 
This diagram yields the large contribution 
\begin{equation}
\delta L_{q,A}=0.203\, \tilde\alpha_s\,,
\label{modelDez}
\end{equation}
where the subscript $A$ stands for the gauge field interaction term.
The total correction to the quark orbital momentum is then given by
$\delta L^{\text{GI}}_q=\delta L_{q}+\delta L_{q,A}$. 
The diagram in Fig.~\ref{OGEgauge} also contributes to the gauge invariant $L^{\text{GI}}_{u-d}$ 
and shifts it by
\begin{equation}
\delta L_{u-d,A}=-\frac{1}{3}\delta L_{q,A}.
\end{equation}
Furthermore, from the gauge-invariant spin sum rule Eq.~(\ref{spin operators2}),  we conclude that the contribution from the total gluon angular momentum equals
\begin{equation}
\label{Jg}
J^{\rm{GI}}_{g}=- \delta L_{q,A}.
\end{equation}
We notice that the corrections (\ref{modelDez})-(\ref{Jg}) are much larger than the known one-gluon exchange contributions from the diagrams of Fig.~\ref{OGE} given in Eqs.~(\ref{resultsDez}),~(\ref{resultsDez2}). In particular, $\delta L_{q,A}$ dominates $L_q^{\rm{MIT}}$ for the chosen parameters.
The MIT bag model results for the gauge invariant decomposition of the nucleon spin are summarized in Table $\ref{modelgauge}$,
together with the combined results, including relativistic effects plus one-gluon exchange corrections plus corrections 
from the pion cloud.

To conclude this section, we emphasize that a direct
calculation of $J^{\rm{GI}}_g = {\langle P+| \int d^3r[\vec r\times(\vec E\times
\vec B)]_3|P+\rangle }$ (i.e. not invoking the spin sum rule) in the framework of
the model requires a careful treatment of the boundary conditions 
for the color electric fields.
The boundary conditions $\hat r\cdot \vec E|_{r=R}$ cannot be fulfilled for
the color electric fields,
in the way described in \cite{Wroldsen:1983ef}. This leads to a non-vanishing
surface term in the calculation
of $J^{\rm{GI}}_g$ and, therefore, potentially to inconsistencies with respect to
the spin sum rule.
A calculation with color electric fields that obey the boundary conditions, as given in \cite{Viollier:1983wy}, turns
out to be significantly more involved and will not
be described in this work.
Note, however, that $\delta L_{q,A}$ is not affected by such complications since the corresponding operator does not involve color electric fields.
It is therefore legitimate to extract the corresponding gluon angular
momentum from $J^{\rm{GI}}_g=1/2-J^{\rm{GI}}_q=-\delta L_{q,A}$. 

When $J^{\rm{GI}}_g$ is calculated directly, using the ``wrong'' color electric fields, it spoils the spin sum rule. Actually the direct evaluation of $J^{\rm{GI}}_g$ can be used to check our result for $\delta L_{q,A}$. Consider the decomposition 
\begin{eqnarray}
\int d^3x\,\vec x \times (\vec E \times \vec B)&=&\int d^3x\, \vec E\times \vec A+\int d^3x\,E^i (\vec x \times \nabla)A^i\nonumber\\
&&\hspace{-30mm}-\int d^3x \,g \psi^\dagger (\vec x \times \vec A)\psi-\int d^3x\nabla^j[E^j(\vec x \times \vec A)].
\end{eqnarray}
The left hand side corresponds to $J^{\rm{GI}}_g$, the right hand side to $\Delta G+L_g-\delta L_{q,A}$ supplemented by a surface term, $-\int d^3x\nabla^j[E^j(\vec x \times \vec A)]$. This surface terms vanishes in the free field theory but in our model calculation this is not the case. Therefore, the total gluon angular momentum calculated through the spin sum rule equals ${\int d^3x\,\vec x \times (\vec E \times \vec B)}+\int d^3x\nabla^j[E^j(\vec x \times \vec A)]$, which indeed can be confirmed by a direct calculation. In fact, the OGE corrections to $\Delta G$ and $L_g$ cancel each other.

\section{Discussion and summary}
\label{Discussion}
The present study has been motivated by the observation of an apparent contradiction between quark orbital angular momentum contributions, $L_q$ and $L_{u-d}$, calculated in models and derived from lattice QCD computations. At the same time, 
model and lattice QCD results for the quark spin contributions $\Delta \Sigma$ and $g_A=\Delta \Sigma_{u-d}$ are reasonably consistent once pion cloud and gluon exchange effects are incorporated in the model \cite{Myhrer:2007cf}. 
When comparing the two approaches, it is essential to note that all spin observables, except $g_A$, are scale (and scheme) dependent quantities. 
Since the model scales are typically low, $\mu^2\sim0.1-0.3\,\rm{GeV}^2$, a careful study of the scale evolution is necessary.
In contrast to previous studies \cite{Thomas:2008ga}, 
our investigations are based on a ``downwards'' evolution of the lattice results, starting at higher scales
where a perturbative treatment appears safe, to low (model) scales where higher-order 
effects must be taken into accout. 
By considering the evolution at LO, NLO and NNLO, together with the possibility that the (non-perturbative) 
strong coupling saturates at very low scales, this approach allows us to perform a meaningful comparison of the lattice 
results and the spin contributions obtained in different model approaches.

With the comparatively low value of $\Lambda^{\rm{LO,\,n_\textsl{F}=3}}_{\rm{QCD}}= 0.148$ $\rm{GeV}$ obtained from the flavor matching procedure, the LO evolution is rather flat for all observables down to scales of $\mu^2=0.1\, \rm{GeV}^2$. At this stage it is not possible to resolve the aforementioned contradiction between phenomenological or lattice results and the model results from Table \ref{model}. This is in contrast to the observation in Ref. \cite{Thomas:2008ga}.

Extending the evolution equations to NLO and NNLO, one finds a strong $\mu^2$-dependence at typical model scales. Even if the matching scale $\mu^2_{\rm{model}}$ is too small to draw 
quantitative conclusions, the trends are indicative. Employing different values for $\Lambda^{\rm{NLO,n_\textsl{F}=3}}_{\rm{QCD}}$ and $\Lambda^{\rm{NNLO,n_\textsl{F}=3}}_{\rm{QCD}}$ as obtained from
flavor matching, the NLO and NNLO results for $J_g$, as well as for $L_{u-d}$, show very good overlap even at the lowest scales. 
For $\Delta \Sigma$ and $L_q$, the results at NLO and NNLO are still quantitatively comparable in magnitude down to scales of about 
$\mu^2\sim0.3 \,\rm{GeV}^2$, but show larger deviations as $\mu^2\rightarrow0.1\rm{GeV}^2$.
In the region of such low scales where the evolution effects become strong, there is indeed an overlap 
with typical model results for the individual observables, as given in Table \ref{model} and \ref{modelgauge}.

Incidentally, the NLO evolved lattice results for $\Delta \Sigma$, $L_q$ and $L_{u-d}$ 
turn out to be reasonably close to the original MIT bag values at a scale of $\mu^2_{\rm{model}}\sim 0.28\,\rm{GeV}^2$ where the total contribution from the gluons vanishes, $J_g(\mu^2_{\rm{model}})=1/2-L_{q}-\Delta \Sigma/2\simeq 0$. 
As the model calculations are improved, however, this apparent consistency deteriorates. Inclusion of the pion cloud in the CBM lowers $\Delta \Sigma$ and increases $L_q$ significantly. The inclusion of the phenomenologically center of mass correction, which is not based on solid theory, would in part restore the agreement. On the other hand, inclusion of further gluon exchange corrections as in Table \ref{modelgauge} would make the matching with the evolved lattice data at a common low scale progressively more difficult. 
For example, while model improvements lead to a significantly larger $L^{GI}_{q}$,
thereby implying a lower matching scale, the corresponding results for the isovector $L^{GI}_{u-d}$ become successively smaller, which
would in turn require a matching at an increasingly higher scale.

It might seem that the scale dependence of $\Delta \Sigma$ implies a large gluon spin fraction $\Delta G$ at the scales of polarized DIS, larger in magnitude than the constraints provided by the HERMES, COMPASS and RHIC measurements \cite{deFlorian:2009vb,:2009ey,:2010um,Abelev:2007vt}. As a test we performed the NLO evolution downward starting from $\Delta G=0$ at $\mu^2=4\,\rm{GeV}^2$. One then finds $\Delta G \sim -0.4$ around the scale where $J_g$ vanishes.

Concerning the lattice calculations, one source of systematic uncertainty can be eliminated by studying isovector quantities such as $L_{u-d}$ for which disconnected diagrams, not taken into account in Ref.\cite{Bratt:2010jn}, cancel out. From the model investigations (see Tables \ref{model}, \ref{modelgauge}) one expects $L_{u-d}^{(GI)}$ in the range of $0.1-0.3$.\footnote{Note, however, that a recent calculation using a chiral quark soliton model gives a negative $L_{u-d}$ even at low scales \cite{Wakamatsu:2009gx}.} In contrast, the lattice results start negative at $\mu^2\sim 4\, \rm{GeV}^2$. 
We find that the downward QCD evolution does indeed predict the appropriate change of sign (see Fig.~\ref{evolutionall}d) at NLO and NNLO.
Other systematic uncertainties on the lattice side, for example those related to lattice operator renormalization issues, would affect the normalization of $L_{u-d}$ but would not change this picture significantly. 
As already noted above, and in contrast to the singlet $L_q$, inclusion of explicit gluon degrees of freedom in the properly gauge invariant treatment of the quark orbital momentum operator leads to a reduction of $L_{u-d}$ at model scales (see Tables \ref{model}, \ref{modelgauge}) and moves this quantity closer to the extrapolated lattice QCD results. The sign change of $L_{u-d}$ can be traced in detail by examining the crossing of $L_u$ and $L_d$ as shown in Fig.~\ref{evolutionLuLd}.

In summary, our analysis underlines the difficulty of a simultaneous, quantitative 
understanding of model calculations and lattice QCD results for the decomposition of the nucleon's spin into the angular momenta of the constituents.
While the perturbative corrections from NLO to NNLO for the evolved lattice results are at a tolerable level,
the broad bands we obtain for a non-perturbative, saturated $\alpha_s$ indicate potentially large systematic 
uncertainties in the evolution at very low scales. 
On the side of the model calculation, we find that the effects from different types of improvements
(related to the pion cloud effects, manifestly gauge invariant OAM operators, one-gluon exchange corrections)
tend to make it increasingly difficult to find a common low matching scale where at least a semi-quantitative agreement
with the evolved lattice results can be achieved for all of the different spin observables.
Conversely, this indicates that it will be difficult to arrive at quantitatively reliable predictions from model calculations starting at scales smaller than $\mu^2\sim 0.3\,\rm{GeV}^2$ and evolving upward to scales accessible in experiments and related QCD phenomenology.
We stress that this observation is not in contradiction with the phenomenologically very successful approach of
``dynamically generated" (unpolarized) parton distributions (PDF) \cite{Gluck:1994uf,Gluck:1998xa,JimenezDelgado:2008hf}, 
where specific ans\"atze for the $x$-dependent PDFs 
are evolved from a very low initial scale, $\mu_0<1\,\rm{GeV}$, to higher scales.
The latter approach necessarily involves adjustable parameters (for each type of PDF) 
to achieve a fully quantitative describtion of the experimental DIS data at scales 
$Q^2=\mu^2>1\,\rm{GeV}^2$ and over a wide range of the momentum fractions $x$.

A possible exception concerning the previous critical assessment is the isovector orbital angular momentum combination $L_{u-d}$ for which systematic lattice errors are minimal. This quantity displays generic behavior with a sign change as it evolves from lattice QCD to low scales, in accordance with the model expectations listed in Tables \ref{model} and \ref{modelgauge}. The stability of this evolution, as one proceeds from NLO to NNLO, becomes particularly apparent when plotted as a function of $\alpha_s$ and compared with the corresponding evolution of the average momentum fraction, $\langle x\rangle_{u-d}$.

\section*{Acknowledgments}
It is a pleasure to thank Werner Vogelsang, Markus Diehl, and Marek Karliner for helpful discussions. This work has been partially supported by BMBF, GSI, and by the DFG cluster of excellence ``Origin and Structure of the Universe''. MA and PH gratefully acknowledge the support by the Emmy-Noether program of the DFG. PH is supported by the SFB/TRR-55 of the DFG. EMH thanks Wolfram Weise for his hospitality and the A.~v.~Humboldt Foundation for a (month-long) stay at the Technische Universit\"at M\"unchen.

\end{document}